\documentclass[conference, 10pt]{IEEEtran}

\usepackage{amsmath}
\usepackage{graphicx}
\usepackage{color}
\usepackage{placeins}
\usepackage{float}
\usepackage{tabularx,colortbl}
\usepackage{psfrag}
\usepackage{epsfig}
\usepackage{subfigure}
\usepackage{multicol}
\usepackage{cite}
\usepackage[usenames,dvipsnames]{xcolor}
\usepackage{jabbrv}

\begin{document}

\title{PML Inspired Transparent Metamaterials}

\author{\IEEEauthorblockN{Nima Chamanara and Christophe Caloz}
\IEEEauthorblockA{Polytechnique Montr\'{e}al,\\ Montr\'{e}al, Qu\'{e}bec H3T 1J4, Canada. \\e-mail: nima.chamanara@polymtl.ca}
}

\maketitle

\begin{abstract}
%\boldmath
Perfectly transparent metamaterial structures of arbitrary shapes, constructed from coordinate stretching and contractions, are presented. Coordinate stretching has been used for 2 decades in perfectly matched layers (PMLs) to electromagnetically simulate infinite domains in numerical techniques, but this concept is applied here for the first time to realize a physical transmission medium. The transparent medium does not scatter electromagnetic waves, i.e. it is reflection-less for all incidence angles and all excitation frequencies. It may be implemented in the form of metasurfaces and will clearly find a myriad of applications if it can be efficiently manufactured.
\end{abstract}

\begin{IEEEkeywords}
Metamaterial, perfect matched layer (PML), coordinate transformation, transformation optics, electromagnetic transparency.
\end{IEEEkeywords}

\IEEEpeerreviewmaketitle

\section{Introduction}

Perfectly matched layers (PMLs) are indispensable tools in the numerical simulation of electromagnetic problems. They absorb incident waves and effectively truncate infinite domains~\cite{taflove2005computational, margengo1999optimum, harari2000analytical}. A PML is essentially an artificial wideband and angle-independent lossy absorbing (non-reflective) layer. Since its introduction in 1994~\cite{berenger1994perfectly}, different variants with specific numerical formulations have been developed, including the split-field PML, the convolutional PML and the uniaxial PML~\cite{berenger2007perfectly}.

In general, absorbing layers may be interpreted as areas of space with stretched coordinates according to a complex stretch factor~\cite{berenger2007perfectly}. As Maxwell equations are invariant under coordinate transformation, such stretching leads to a new electromagnetic solution, corresponding to new material parameters in the initial (non-stretched) system. This material can be obtained through transformation electromagnetics or transformation optics techniques~\cite{pendry2006controlling, pendry2012transformation}. For a PML in Cartesian coordinates, coordinate stretching leads to an anisotropic uniaxial medium.

PMLs are essentially artificial and purely numerical media. However, the particular case of the uniaxial PML represents in fact an actual physical medium. Inspired by this consideration, we apply here coordinate stretching and contraction to construct metamaterials that do not scatter electromagnetic waves whatever their shape is. Given their fundamental nature, such metamaterials may find a wide range of applications in daily life.

The organization of the paper is as follows. Section~\ref{sec:pml_transem} explains the connection of the PML concept to coordinate stretching and transformation electromagnetics. Section~\ref{sec:pml_transp_structs} describes  PML-based perfectly transparent structures and applications. Conclusions are given in Sec.~\ref{sec:concl}.

\vspace{3mm}

\section{PML as coordinate transformation}\label{sec:pml_transem}

\begin{figure}[ht!]
\subfigure[]{ \label{fig:plwave_coord_stretch_no}
\psfrag{x}[c][c][0.9]{$x$}
\psfrag{y}[c][c][0.9]{$y$}
\includegraphics[width=0.3\columnwidth]{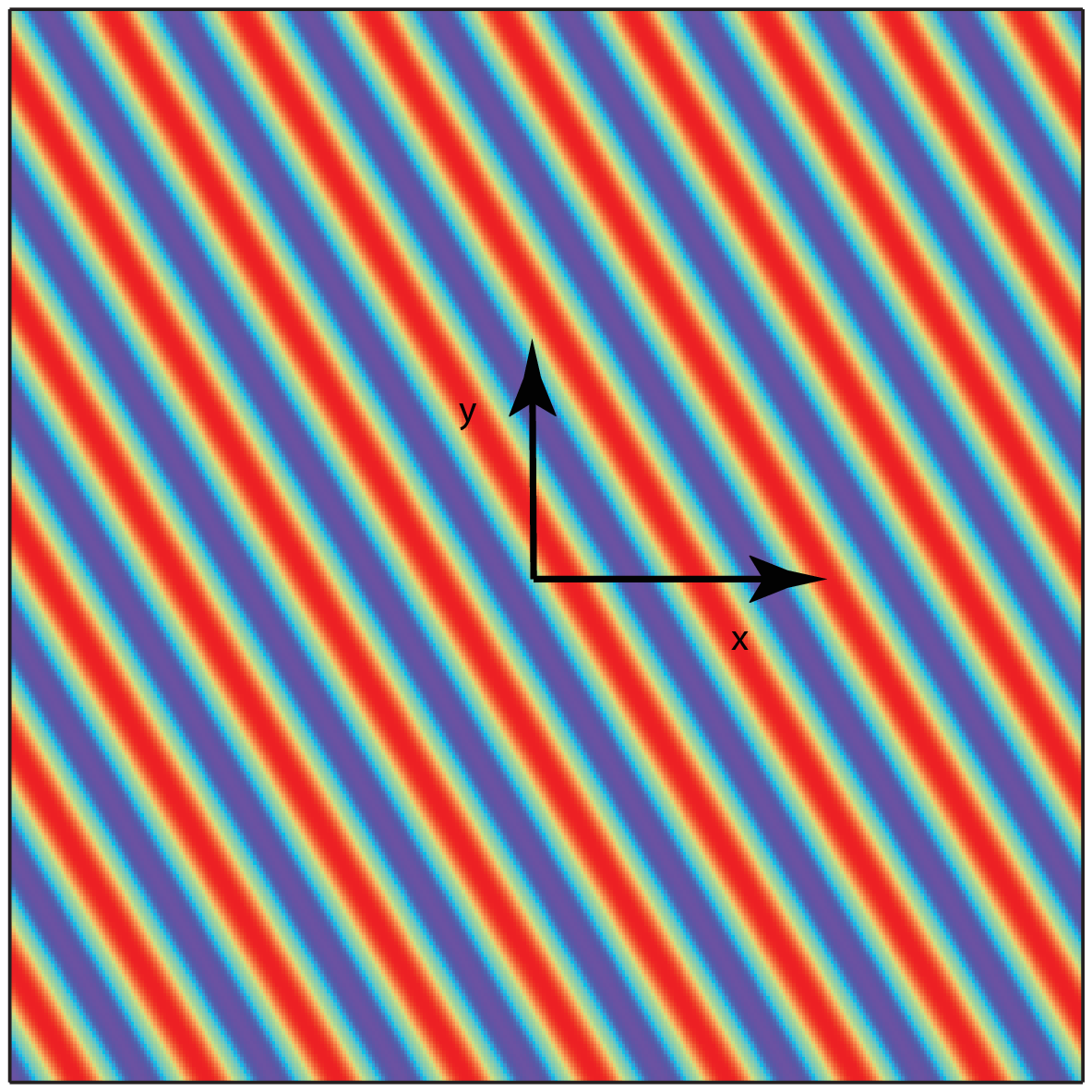}}
\subfigure[]{ \label{fig:plwave_coord_stretch_real}
\includegraphics[width=0.3\columnwidth]{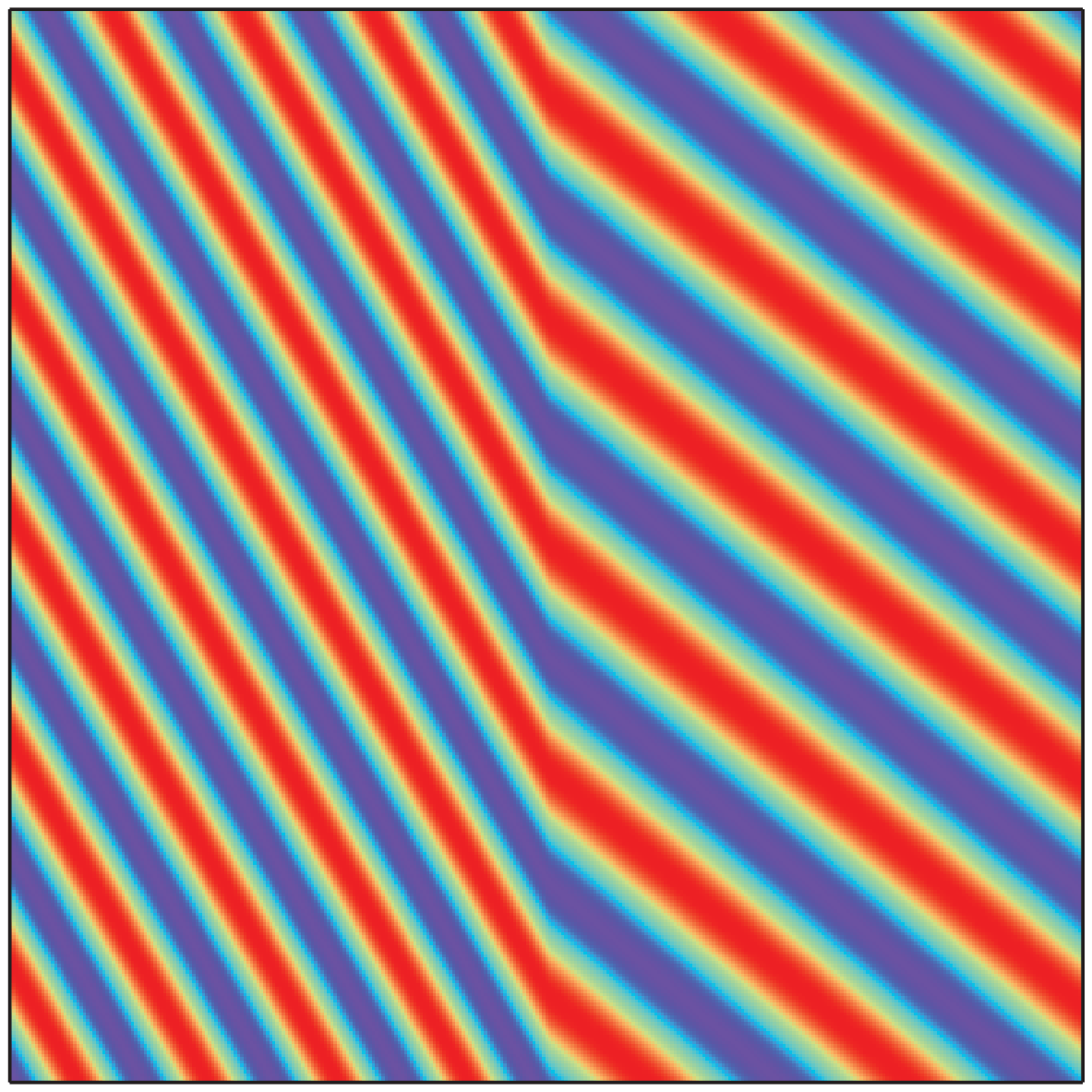}}
\subfigure[]{ \label{fig:plwave_coord_stretch_complex}
\includegraphics[width=0.3\columnwidth]{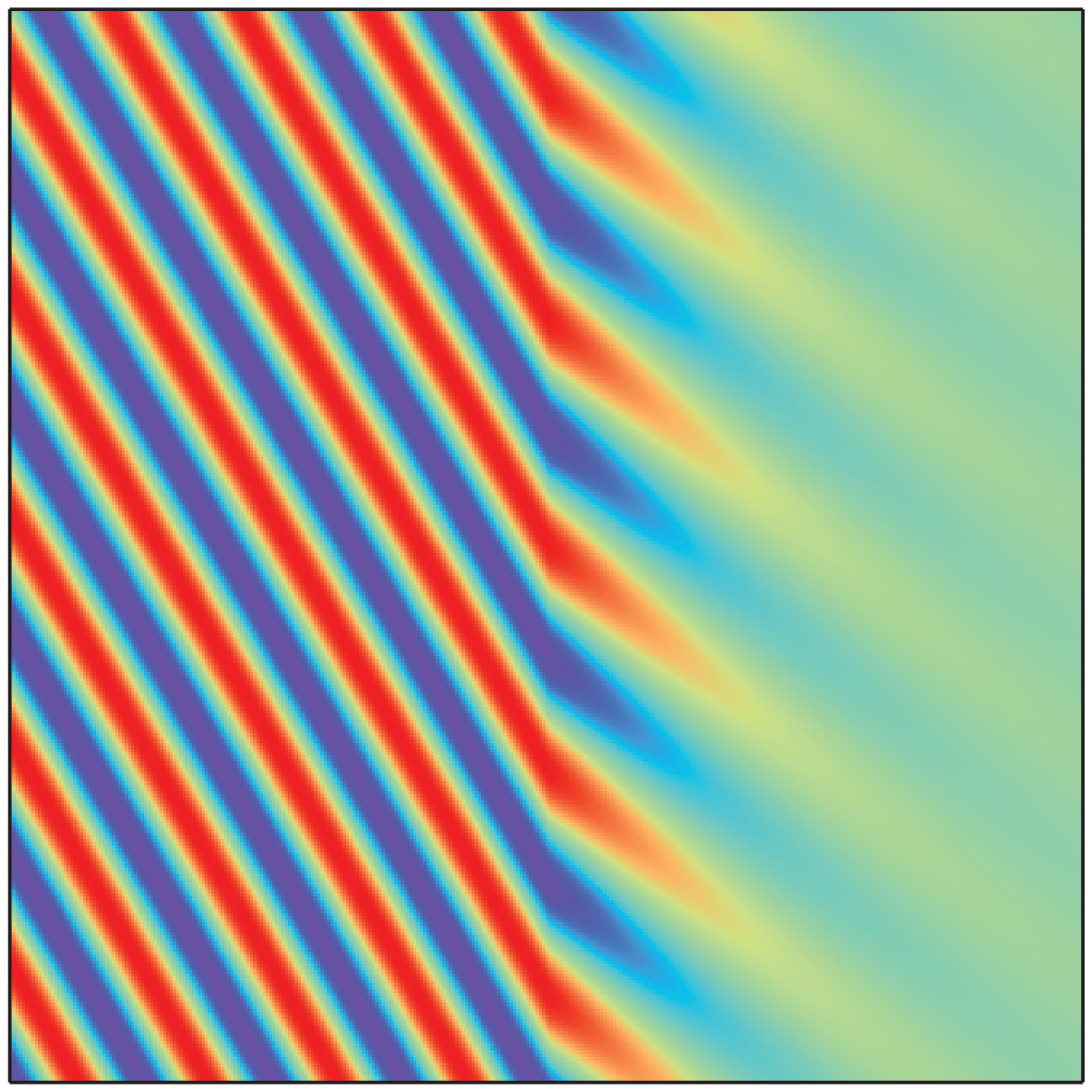}}
\caption{Demonstration of coordinate stretching. (a)~Oblique free space plane wave. (b)~Region $x>0$ stretched by a real stretch factor $s_x$. (c)~Region $x>0$ stretched by a complex stretch factor $s_x$.}
\label{fig:plwave_coord_stretch}
\end{figure}

A PML can be described as a coordinate transformation, more specifically a coordinate stretching with a complex stretch factor, as illustrated in Fig.~\ref{fig:plwave_coord_stretch}.

Imagine a solution of Maxwell equations, such as an oblique plane wave $\mathbf{E} = \mathbf{E_0}\exp[-j(k_x x + k_y y)]$, shown in Fig.~\ref{fig:plwave_coord_stretch_no}. If the $x$~axis is stretched by a factor $s_x$, as shown in Fig.~\ref{fig:plwave_coord_stretch_real} for $s_x$ real and in Fig.~\ref{fig:plwave_coord_stretch_complex} for $s_x$ complex, this plane wave is transformed into the new plane wave $\mathbf{E} = \mathbf{E_0} \exp[-j(s_x k_x x + k_y y)]$, which is not a free space solution of Maxwell equations anymore. However, according to transformation electromagnetics, if the transformed region is filled with a medium characterized by the relative permittivity and permeability tensors~\cite{pendry2012transformation}

\begin{equation}
\bar{\bar{\epsilon}}_r=\bar{\bar{\mu}}_r=\frac{\mathbf{A A^T}}{\det(\mathbf{A})},
\end{equation}

\noindent where $\mathbf{A}$ is the Jacobian matrix of the transformation, the transformed fields remain a valid solution of Maxwell equations in the new medium. For a general stretching in the Cartesian system with stretch factors $s_x$, $s_y$ and $s_z$ along the $x$, $y$ and $z$~axis, respectively, the medium parameters are found as

\begin{equation}\label{eq:eps_mu_sxyz}
\bar{\bar{\epsilon}}_r=\bar{\bar{\mu}}_r=\left[\begin{matrix}\frac{s_{x}}{s_{y} s_{z}} & 0 & 0\\0 & \frac{s_{y}}{s_{x} s_{z}} & 0\\0 & 0 & \frac{s_{z}}{s_{x} s_{y}}\end{matrix}\right].
\end{equation}

\noindent
Such a medium is perfectly matched to free space for all angles of incidence. For complex stretch factors, the medium parameters~\eqref{eq:eps_mu_sxyz} describe the uniaxial PML \cite{berenger2007perfectly}, which fully absorbs the incident waves.

\vspace{3mm}

\section{PML Inspired Transparent Metamaterial} \label{sec:pml_transp_structs}

Historically, the stretch factor has been a complex number, in order to attenuate the waves penetrating into the PML medium, and the PML layer has been truncated with a perfect electric conductor (PEC) wall. Such a choice suits the purpose of an absorbing zone in an electromagnetic simulation. However, we propose here to apply this transformation in a very different fashion, to devise a novel metamaterial.
  
\begin{figure}[ht!]
\subfigure[]{
\psfrag{m}[c][c][0.9]{glass}
\psfrag{i}[c][c][0.9]{$\mathbf{E}^I$}
\psfrag{r}[c][c][0.9]{$\mathbf{E}^R$}
\psfrag{t}[l][c][0.9]{$\mathbf{E}^T$}
\includegraphics[width=0.45\columnwidth]{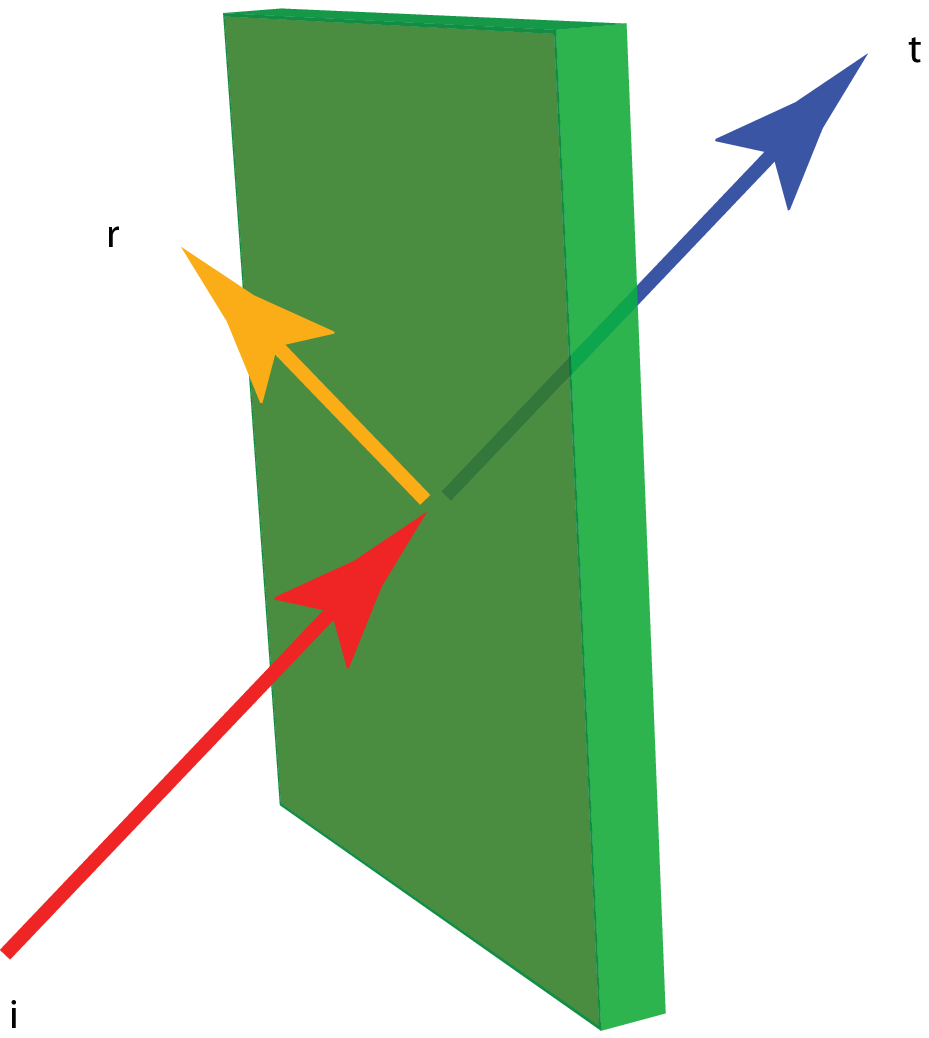}}
\subfigure[]{
\psfrag{m}[c][c][0.9]{perfect glass}
\psfrag{i}[c][c][0.9]{$\mathbf{E}^I$}
\psfrag{t}[l][c][0.9]{$\mathbf{E}^T$}
\includegraphics[width=0.45\columnwidth]{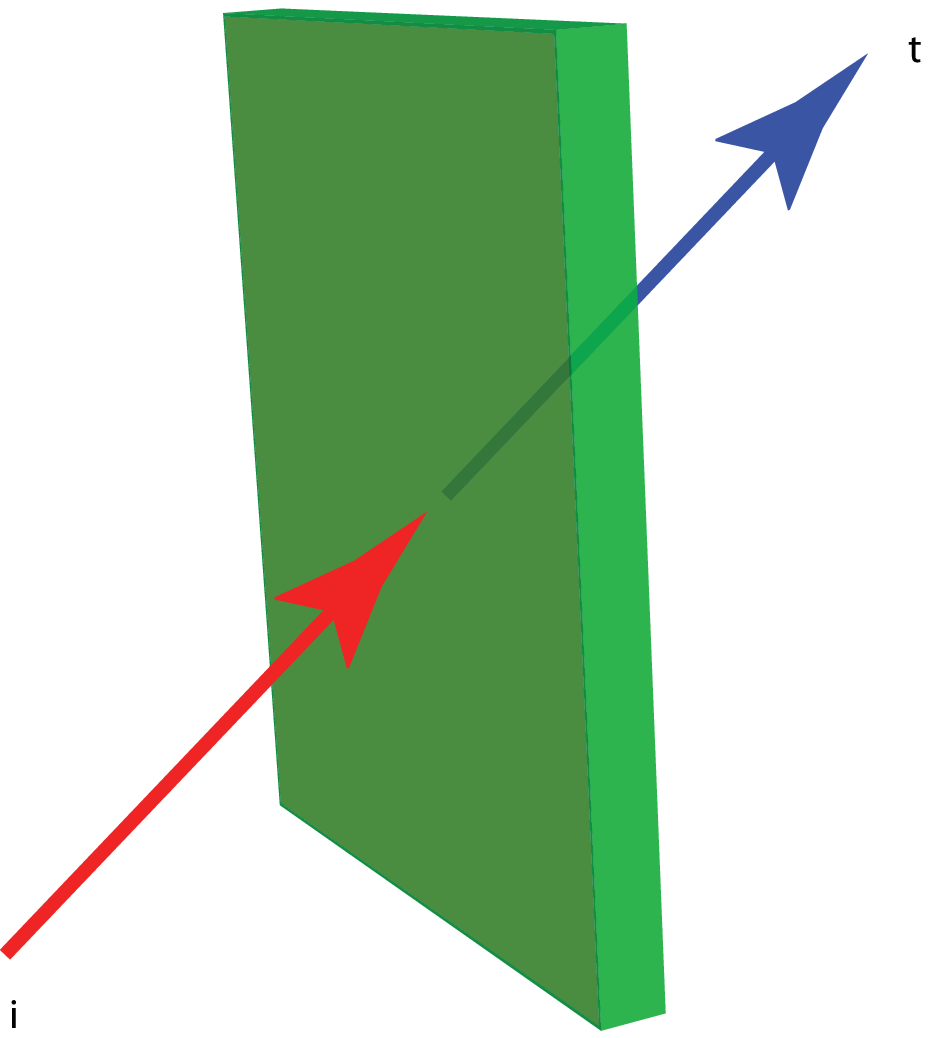}}
\caption{Concept of ``perfect glass'' or slab. (a)~Ordinary glass and slabs generally suffers from reflection. (b)~A ``perfect glass'' or slab is perfectly transparent for all angles of incidence and all excitation frequencies.}
\label{fig:mtm-glass-rect}
\end{figure}

We realized that a finite-thickness area of space that has been coordinate-stretched with a \emph{real stretch factor} along a direction (direction $x$ in~Fig.~\ref{fig:plwave_coord_stretch}) describes in reality a perfectly transparent slab. Specifically, this slab is perfectly reflection-less for all incidence angles and all excitation frequencies. It would correspond to ``perfect glass'' in daily life applications where glass reflections are generally undesirable. This ``perfect glass'' concept is illustrated in Fig.~\ref{fig:mtm-glass-rect}. In the optical regime, applications would include laptops, tablets, cell phones, TV screens, window glass, cockpit glass, etc. In the radio regime, applications may include microwave-wise transparent walls for wireless communications or stealth structures for defense.

\begin{figure}[ht!]
\subfigure[]{
\psfrag{m}[c][c][0.9]{glass}
\psfrag{i}[c][c][0.9]{$\mathbf{E}^I$}
\psfrag{r}[c][c][0.9]{$\mathbf{E}^R$}
\psfrag{t}[c][c][0.9]{$\mathbf{E}^T$}
\includegraphics[width=0.45\columnwidth]{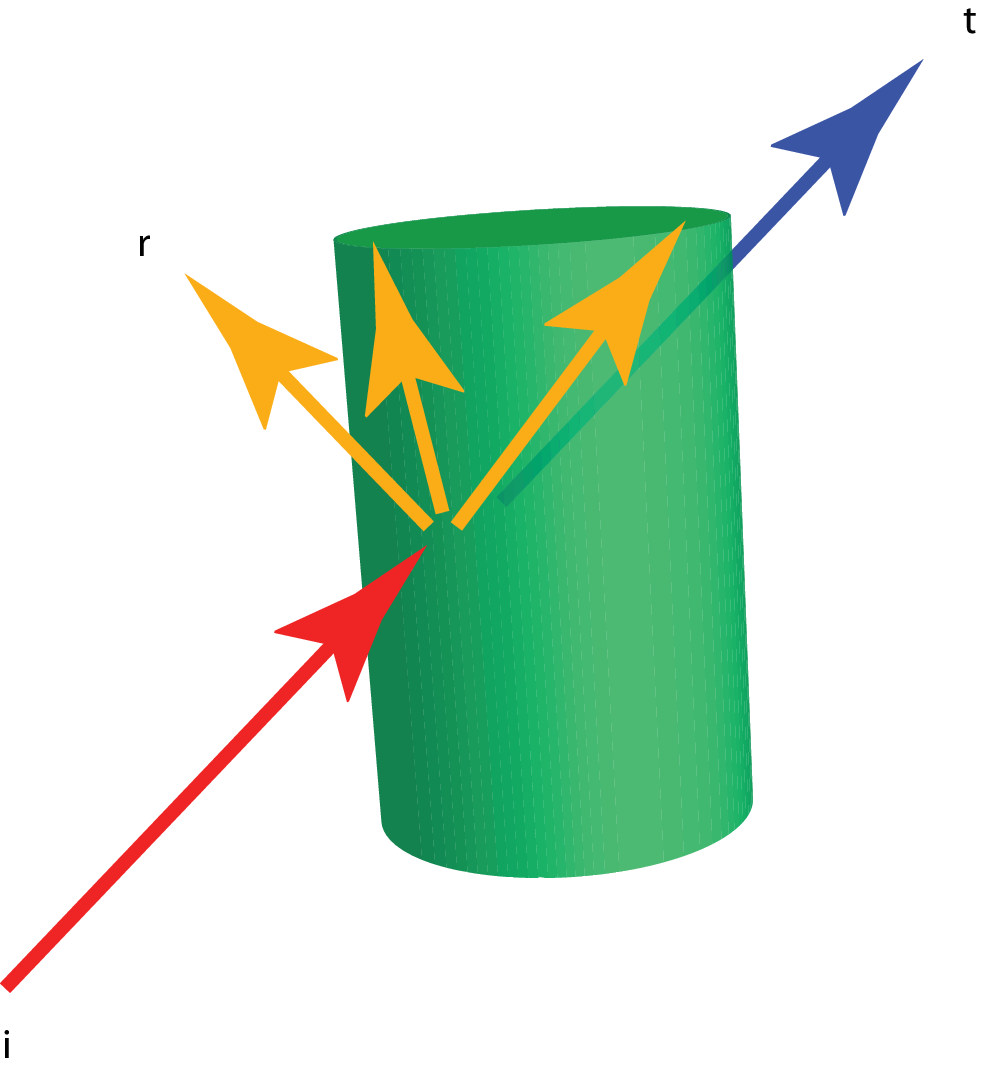}}
\subfigure[]{
\psfrag{m}[c][c][0.9]{perfect glass}
\psfrag{i}[c][c][0.9]{$\mathbf{E}^I$}
\psfrag{t}[c][c][0.9]{$\mathbf{E}^T$}
\includegraphics[width=0.45\columnwidth]{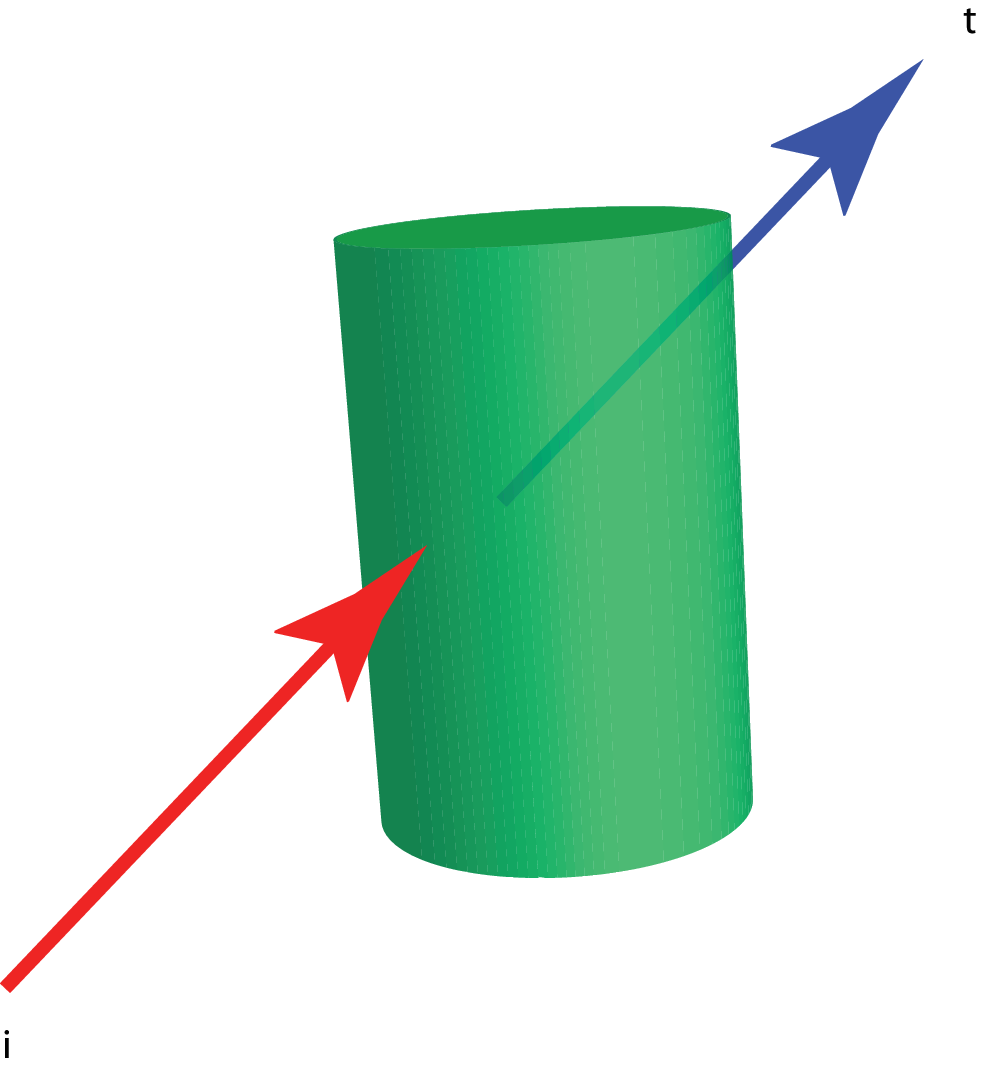}}
\caption{Same as in Fig.~\ref{fig:mtm-glass-rect} but for a cylindrical rod shape. With proper stretching and contraction of space, perfect transparent metamaterial structures of different shapes can be produced.}
\label{fig:mtm-glass-cyli}
\end{figure}

Using proper stretching and contraction of space in the appropriate coordinate system, it is in principle possible to extend the concept of perfectly transparent materials to arbitrary shapes. An illustration of such an extension is a perfectly transparent cylindrical rod, shown in Fig.~\ref{fig:mtm-glass-cyli}. As the slab, this structure does not scatter any wave, whatever its angle of incidence or frequency.

It should be noted that, although invisible, the proposed metamaterials are \emph{not} invisibility cloaks~\cite{pendry2006controlling}. An object inside an invisibility cloak does not interfere with the incident light and therefore remains imperceptible to an outside observer. In contrast, an object inside or in the vicinity of a perfect transparent material scatters the incident light as if the surrounding material would not exist. Such a medium represents perfect test bed for scattering measurements, as it does not alter the field scattered from the objects under measurement.

\vspace{3mm}

\section{Conclusions}\label{sec:concl}

Perfectly transparent metamaterial structures with arbitrary shapes constructed from coordinate stretching and contraction have been presented. Such coordinate stretching has been historically used for perfectly matched layers (PMLs) in the electromagnetic simulation of infinite domains. The proposed transparent structures do not scatter incident electromagnetic waves; they are reflection-less for all angles of incidence and all excitation frequencies. Such structures may find numerous applications in daily life.

\vspace{5mm}

\bibliographystyle{jabbrv_ieeetr}
\bibliography{ReferenceList2}

\end{document}